\documentclass[preprint,12pt]{elsarticle}
\usepackage[top=1cm]{geometry}

\usepackage{amssymb}
\usepackage{amsmath}
\usepackage{lineno}

\usepackage{hyperref}
\usepackage{caption}
\usepackage{subcaption}
\begin{document}

\begin{frontmatter}



\title{The homeostatic dynamics of feeding behaviour identify novel mechanisms of anorectic agents}


\author[1,3]{Thomas M McGrath}
\author[2]{Eleanor Spreckley}
\author[2]{Aina Fernandez Rodriguez}
\author[4]{Carlo Viscomi}
\author[2]{Amin Alamshah}
\author[2]{Elina Akalestou}
\author[2]{Kevin G Murphy\corref{cor2}}
\author[1,3]{Nick S Jones\corref{cor1}}
\address[1]{Department of Mathematics, Imperial College London}
\address[2]{Section of Endocrinology and Investigative Medicine, Imperial College London}
\address[3]{EPSRC Centre for the Mathematics of Precision Healthcare, Imperial College London}
\address[4]{MRC Mitochondrial Biology Unit, University of Cambridge}
\cortext[cor1]{Corresponding author/lead contact - nick.jones@imperial.ac.uk}
\cortext[cor2]{Co-corresponding author - k.g.murphy@imperial.ac.uk}

\begin{abstract}
Better understanding of feeding behaviour will be vital in reducing obesity and metabolic syndrome, but we lack a standard model that captures the complexity of feeding behaviour. We construct an accurate stochastic model of rodent feeding at the bout level in order to perform quantitative behavioural analysis. Analysing the different effects on feeding behaviour of PYY\textsubscript{3-36}, lithium chloride, GLP-1 and leptin shows the precise behavioural changes caused by each anorectic agent. Our analysis demonstrates that the changes in feeding behaviour evoked by the anorectic agents investigated not mimic satiety. In the \textit{ad libitum} fed state during the light period, meal initiation is governed by complete stomach emptying, whereas in all other conditions there is a graduated response. We show how robust homeostatic control of feeding thwarts attempts to reduce food intake, and how this might be overcome. \textit{In silico} experiments suggest that introducing a minimum intermeal interval or modulating gastric emptying can be as effective as anorectic drug administration.
\end{abstract}


\end{frontmatter}

%\linenumbers

\section{Introduction}
The current obesity epidemic has driven significant interest in developing novel therapies to reduce food intake. However, the homeostatic nature of energy balance makes feeding a difficult process to control; manipulating one aspect of feeding behaviour can result in compensatory changes in other aspects, such that it is possibly to radically restructure feeding without altering total food intake over a chronic period \citep{moran2004gastrointestinal,ellacott2010assessment,campos2016parabrachial}. A more detailed understanding of how feeding behaviour is structured might allow understanding of which aspects may most promisingly be targeted to prevent or treat obesity. One natural avenue of investigation is to look in greater detail at feeding behaviour: how feeding plays out through time at the individual level, paying attention to behaviour at finer-grained temporal resolution. A natural resolution is the level of individual feeding bouts and their organisation into meals \citep{tolkamp1998satiety,tolkamp2011temporal}, which is conventionally referred to as microstructure analysis \citep{davis1989microstructure}. In this paper we have analysed the microstructure of feeding behaviour by developing a mathematical model of food intake which we then use to carry out a detailed examination of feeding behaviour using bout level data, and the way in which the gut governs the dynamics of feeding. This model provides novel insight into the mechanisms by which anorectic agents act to alter feeding behaviour, and how food intake changes with nutritional status and photoperiod.
\newline\newline
Modelling homeostatic behaviours such as feeding poses a particular challenge. Although recent advances in automated monitoring technology can create large datasets through quantitative, high-throughput capture and annotation of behaviour \citep{schaefer2012surveillance, gomez2014big, anderson2014toward, egnor2016computational, johnson2016composing, krakauer2017neuroscience}, homeostatic behaviours are driven by the need to regulate a (typically unobserved and continuously-varying) variable such as body temperature or energy balance. Thus we expect the rates of transitions between behavioural states to be strongly modulated by the variable under regulation. Conventional behavioural analysis makes heavy use of the ethogram, which charts the rates of transitions between different behaviours. However when these rates vary this is no longer possible, and new techniques are needed. In this paper we suggest such a technique: a class of models known as piecewise deterministic Markov processes \citep{davis1984piecewise}. These generalise Markov Chains to include exactly the kind of deterministically-varying transition rates required by models of homeostatic behaviour and have found a wide variety of uses in biology \citep{rudnicki2015piecewise, beil2009simulating, azais2014piecewise, bodova2017probabilistic} and physics \citep{faggionato2009non, bressloff2017hamiltonian}, although they have not previously been used to study food intake or energy balance. We use this model class to create a flexible and intuitive stochastic model of feeding behaviour governed by stomach fullness. This allows us to both infer characteristics of feeding behaviour both for individuals and groups of animals, and to generate representative behaviour sequences \textit{in silico} for a previously characterised group or individual. Although the model we introduce in this work only considers the effect of stomach fullness on feeding behaviour and how this is modulated by anorectic agents fasting and the day/night cycle, it can easily be extended to incorporate other factors that are known to be important such as food palatability, energy expenditure, or social factors. This could be achieved either by introducing new dynamic variables (for instance in the case of energy expenditure) or extending the current analysis to include groups in differing conditions. Social factors influencing behaviour have previously been modelled within the mathematical framework we use \cite{bodova2017probabilistic}, demonstrating the effectiveness of this approach. The only previous generative model of feeding known to the authors is \citep{booth1978hunger}, which does not allow for inference from data for either individuals or groups. Previous attempts to characterise the effects of food intake on feeding behaviour have centred on the satiety ratio, which measures how the size of a meal affects the intermeal interval that follows it (see Supplementary Material). We show that the satiety ratio cannot successfully model feeding behaviour as it ignores the effects of earlier meals. Furthermore, it does not define a generative model, preventing \textit{in silico} experimentation. In this paper we undertake a quantitative study of feeding microstructure in rats in a wide variety of conditions with the aim of better understanding how they affect behaviour. We generated data from rats in both photoperiods, both fasted and \textit{ad libitum} fed, and given a range of anorectic drugs. In order to investigate the wider validity of the model we also generated data from \textit{ad libitum} fed mice in both the light and dark periods. This data, which, when combined with our model, allowed us to identify novel mechanisms of behavioural action for anorectic drugs, replicate recent \textit{in vivo} results \textit{in silico}, and investigate how behavioural interventions can combine with anorectic drug administration to robustly reduce food intake.

\section{Results}
\subsection{High-resolution feeding data is naturally captured by a stochastic model}
Recent evidence indicates that calorie content in the stomach rapidly and powerfully controls food intake at least in part by signalling via Agouti-related peptide neurons \citep{beutler2017dynamics,su2017nutritive}. When considered in combination with previous evidence for the key role of the gut in controlling both meal termination and initiation \citep{janssen2011role} this makes the gut a promising candidate for explaining the dynamics of feeding behaviour. The central role of stomach fullness in governing food intake makes it a promising quantity to consider when modelling feeding behaviour. We constructed a model of rodent feeding behaviour that takes into account stomach fullness, summarised in Figure~\ref{fig:model_intro}A, which we then used to determine how the different components of feeding vary in different conditions, and to conduct \textit{in silico} experiments to investigate the effects of behavioural changes. Our model uses the conventional partitioning of feeding into bouts and meals: within a meal, bouts are punctuated by short pauses, with longer intermeal intervals occupying the time between meals. A number of behavioural correlates including the behavioural satiety sequence indicate that this is a meaningful distinction, and that intermeal intervals are not simply unusually long inter-bout pauses \citep{tolkamp1998satiety}. We used bout-level feeding data obtained from the Columbus Instruments Comprehensive Lab Animal Monitoring System (CLAMS), although other systems such as BioDAQ can provide suitable data. This data was then used as input to a simple model of stomach filling and emptying \citep{booth1978hunger} to provide a continuous measurement of stomach fullness through time, as shown in Figure~\ref{fig:model_intro}B and C. Although we have described the parameter $x$ as representing stomach fullness, its effect on behaviour may also reflect the influence of the upper part of the small intestine. The model thus contains two parts: an empirical model of stomach filling and emptying (Figure~\ref{fig:model_intro}C) and a model of transitions between behavioural states (Figure~\ref{fig:model_intro}D-G). The model of stomach filling and emptying is deterministic conditional on knowing the rodent's behavioural state, whereas the behavioural model is stochastic, reflecting the unpredictability of behaviour. The model can be summarised as follows: a feeding bout occurs with stochastic duration and feeding rate. This increases stomach fullness. Following the termination of this feeding bout, a meal termination decision is made which is also stochastic, but depends on stomach fullness. If the meal is not terminated, a short within-meal pause occurs, whose duration is stochastic but independent of stomach fullness. If the meal is terminated, however, then a long intermeal interval occurs, whose duration depends on stomach fullness. The distribution of each stochastic event is controlled by parameters that vary between rats but are fixed for each individual in a group (nutritional state, drug administration, and photoperiod), summarised in Table~\ref{tab:params}. The degree of variation between rats in the same group is controlled by group-level parameters $\mu$ and $\Sigma$, summarised in the same table. These parameters define the `typical' rat in the group, as well as how much rats within the group vary.

\subsection{Our stochastic model can accurately capture rat feeding behaviour in a wide variety of conditions}
We carried out a series of experimental studies in order to obtain behavioural data from rats given low, medium, and high doses of peptide YY\textsubscript{3-36} (PYY\textsubscript{3-36}) in the light and dark period, as well as rats given low and high doses of lithium chloride (LiCl) in the light period and rats given low, medium, and high doses of glucagon-like peptide 1 (GLP-1) and leptin in the dark period (details and doses in Methods). From these experiments we also obtained control data from rats given saline under each of these conditions. All experiments in the light period were carried out following an overnight fast, whereas dark period experiments involved \textit{ad libitum} feeding. In addition we carried out a longer duration feeding study over three days with untreated rats recovering from a fast, providing further data on `natural' behavioural patterns. We further obtained bout-level feeding data from mice in order to determine whether our model had a wider range of applicability. This exploratory analysis, and details of data collection and processing, is detailed in Supplementary Material section 4. We found that with minimal reparametrisation in order to account for the different total food intake in mice we were able to model mouse feeding behaviour and recapitulate a novel tradeoff in ingestive behaviour that we discovered in rats (see below).

\begin{table}
\begin{footnotesize}
\centering
\begin{tabular}{ccp{0.3\textwidth}}
\textbf{Parameter} & \textbf{Description} & \textbf{Comments}\\
\hline
\hline
\textbf{Feeding bout}&&\\
\hline
$\lambda_{F}$&Feeding bout termination rate&Probability of feeding bout ending per unit time - sets mean bout duration\\
$\rho_{F}$&Feeding rate&Rate of feeding in a bout - randomly sampled for each bout\\
$\mu_{F}$&Mean bout feeding rate&Bout feeding rates are normally distributed for each animal - this sets the mean\\
$\sigma_{F}$&Feeding rate variance&Sets the variance of the normal distribution of feeding rates\\
\hline
\textbf{Short within-meal pause}&&\\
\hline
$\lambda_{S}$&Short pause termination rate&Probability of a short within-meal pause ending per unit time - sets mean short pause duration\\
\hline
\textbf{Meal termination}&&\\
\hline
$T_{1}$&Meal termination parameter 1&Makes meal termination probability curve sharper (see Figure~\ref{fig:model_intro}F)\\
$T_{2}$&Meal termination parameter 2&Makes meal termination more stomach-sensitive (see Figure~\ref{fig:model_intro}F)\\
\hline
\textbf{Intermeal interval}&&\\
\hline
$L_{1}$&Intermeal interval parameter 1&Sets stomach-independent component of intermeal interval termination rate (see Figure~\ref{fig:model_intro}G)\\
$L_{2}$&Intermeal interval parameter 2&Sets stomach-dependent component of intermeal interval termination rate (see Figure~\ref{fig:model_intro}G)\\
\hline
\textbf{Group characteristics}&&\\
\hline
$\Theta$&Group mean&Defines the `typical individual' for a group - the mean of the group distribution\\
$\Sigma$&Group covariance matrix&Sets the inter-individual variation of parameters and how correlated different parameters are\\
\hline
\textbf{Other}&&\\
\hline
$k$&Gastric emptying parameter&Sets gastric emptying rate via $\dot{x}=-k\sqrt{x}$ (where $x$ is stomach fullness)
\end{tabular}
\end{footnotesize}
\caption{Summary and description of model parameters. See Figure~\ref{fig:model_intro} for a graphical summary and equations.}\label{tab:params}
\end{table}

\begin{figure}[htp]
    \centering
    \includegraphics[width=0.95\textwidth]{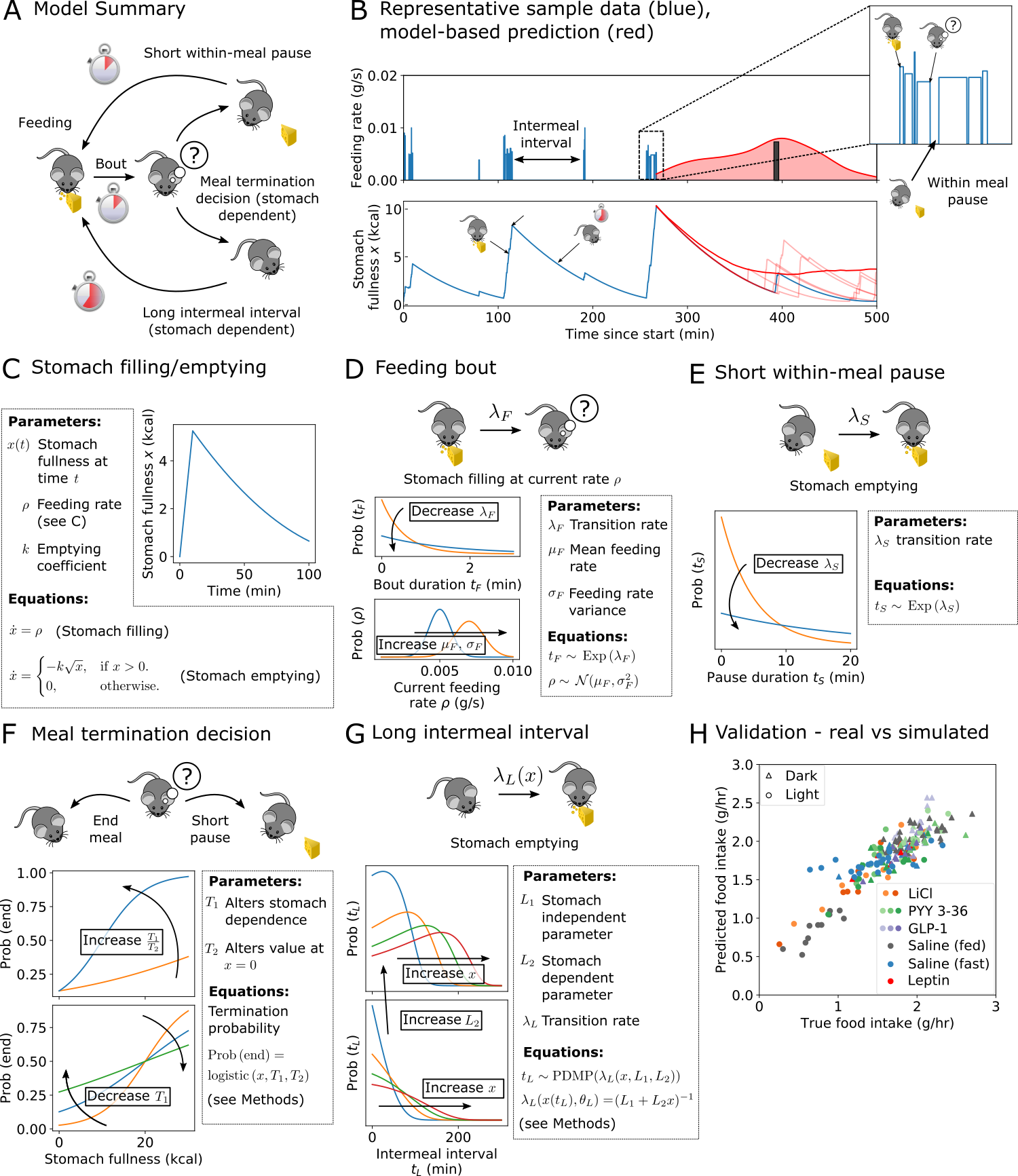}
    \caption{A simple stochastic model of rat feeding that uses a continuously 
    time-varying stomach fullness can accurately recapitulate food intake. \textbf{(A)} Schematic illustration of the model, showing feeding, a short within-meal pause, a long intermeal interval, and the meal termination decision, which relies on stomach fullness $x$. \textbf{(B)} Incorporating stomach fullness into a stochastic model of feeding allows prediction of the intermeal interval. The model predicts a distribution over possible intermeal interval lengths (red curve in top panel, actual next meal time shown as a dark bar). Representative sample of data showing bout-level feeding data (top panel, blue bars) and model-derived stomach fullness (bottom panel, blue curve). Simulated trajectories of stomach fullness are shown in the bottom panel (light red curves) alongside mean fullness (dark red curve). Inset: bout-level feeding data for the meal shown inside the dashed box. \textbf{(C)} Stomach filling occurs linearly at a rate $\rho$ that varies between bouts (see (D)), whereas stomach emptying is nonlinear but has identical dynamics whenever the animal is not feeding. \textbf{(D)} Feeding bout duration is exponentially distributed, and rate $\rho$ is normally distributed with mean $\mu_{F}$ and standard deviation $\sigma_{F}$. \textbf{(E)} Within-meal pauses are typically short compared to intermeal intervals, and are distributed exponentially. Figure shows exponential distribution for representative parameter values. \textbf{(F)} Meal termination decisions are sigmoid in stomach fullness $x$ and are controlled by parameters $T_{1}$ and $T_{2}$. Keeping the product of these parameters fixed changes the effect of the stomach on termination probability. \textbf{(G)} Intermeal intervals are typically long and depend on stomach fullness $x$, which varies over time. Parameter $L_{1}$ controls intermeal interval duration independent of stomach fullness, $L_{2}$ measures the effect of stomach fullness. \textbf{(H)} Our model accurately captures food intake: (posterior) predictive values of normalised food intake are strongly correlated with true values. Marker style indicates anorectic agent (if any) and photoperiod (see legend). Darker colours indicate higher doses, see Table~\ref{tab:drugs} for concentrations.}
    \label{fig:model_intro}
\end{figure}

After fitting the model to the data (see Methods) we validated the model by simulating behaviour to provide predictions on the next meal time, total food consumption, or future stomach fullness, as shown in Figure~\ref{fig:model_intro}B. A subset of the saline and PYY 3-36 data was used in model design, however the model was further validated by comparing fits to unseen saline data with fits to saline data from identical conditions (see Supplementary Material). The long-term organisation of feeding behaviour into meals and bouts occurs on timescales comparable to typical experimental observations: we may only see 5-10 meals for a single rat's time series. This complicates inference if we also want to consider the possibility of inter-individual variation. We used Bayesian hierarchical modelling \citep{gelman2014bayesian,mcelreath2015statistical} to allow pooling of data across individuals in a statistically rigorous way, and to infer the degree of interindividual variation (see Methods). However, in order to verify the model's predictive power, we generated predictions for food intake for each individual by repeatedly Monte Carlo sampling using parameters drawn from individual-level posteriors (see Supplementary Material). Our predictions agreed with observed food intake across a wide range of food intakes from different photoperiods, anorectic drugs, and nutritional states (as shown in Figure~\ref{fig:model_intro}H) in spite of the level of stochasticity inherent in feeding, confirming that our model captured food intake accurately. Code for inference, figure generation, and a set of simple tools for behavioural modelling is available at \url{https://github.com/tomMcGrath/feeding_analysis}, and full details of drug administration doses and protocols are given in Table~\ref{tab:drugs} (see Materials and Methods).

\subsection{Both within-bout and intermeal behaviour determine food intake}
\begin{figure}[htp]
    \centering
    \includegraphics[width=0.95\textwidth]{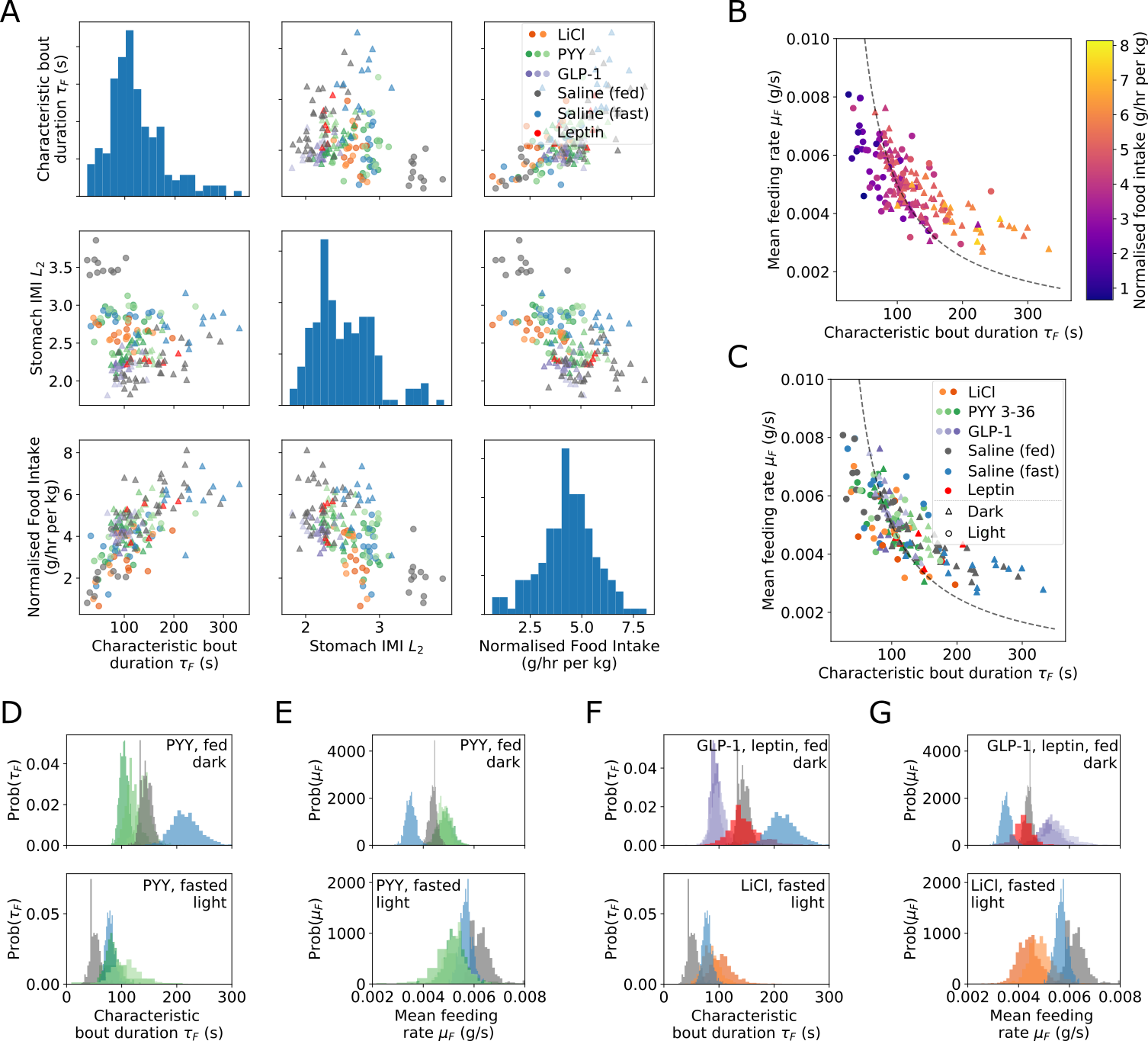}
    \caption{A two parameter behavioural fingerprint, combining characteristic bout duration and intermeal interval parameters accounts for a large proportion of variation both between and within groups. Different anorectic agents drive different patterns of feeding behaviour. \textbf{(A)} \chapter{Two} parameters of the model are strongly informative regarding food intake, but are only weakly correlated with one another: characteristic bout duration $\tau_{F}$ and stomach-dependent intermeal interval parameter $L_{2}$. Scatter matrix of $\tau_{F}$, $L_{2}$ and normalised food intake (diagonal entries show univariate marginals) indicates substantial inter-group variation in $\tau_{F}$ and $L_{2}$ correlated with changes in normalised food intake. \textbf{(B)} Reduced characteristic bout durations are not fully compensated for by increased feeding rate: scatter plot of individual posterior mean values coloured by normalised food intake. Dashed line indicates constant food intake contour. \textbf{(C)} Characteristic bout duration and feeding rate vary both within and between groups. Animals in the dark period tend to have longer, slower bouts. \textbf{(D, E)} PYY\textsubscript{3-36} causes group-level variation in bout time and feeding rate. Colours as in Figure 2C. Animals recovering from an overnight fast show less variation than \textit{ad libitum} fed rats. \textbf{(F, G)} GLP-1 produces a more pronounced effect on both bout duration and rate than PYY\textsubscript{3-36} in \textit{ad libitum} fed rats. Lithium chloride produces a stronger reduction in feeding rate than PYY\textsubscript{3-36} in rats recovering from an overnight fast, but has limited effect on bout duration.}
    \label{fig:fingerprint}
\end{figure}

When each of the model parameters were compared with mean hourly food intake, two stood out as clear determinants of food intake: mean bout duration $\tau_{F} = 1/\lambda_{F}$ and the parameter governing the dependence of the intermeal interval on stomach fullness $L_{2}$, (Figure~\ref{fig:fingerprint}A). Bout duration and stomach-dependent intermeal interval are strongly related to food intake: decreasing the former leads to shorter meals, and increasing the latter means that stomach fullness strongly suppresses meal initiation, leading to long intermeal intervals in most circumstances. These two important parameters are found to be only weakly correlated with one another, but are both strongly correlated with food intake (Figure~\ref{fig:fingerprint}A), making them an informative way to determine what behavioural changes may have led to an observed change in food intake. The stomach-dependent intermeal interval parameter $L_{2}$ shows correlation with food intake across but not within groups, suggesting that although different conditions and anorectic agents can cause substantial changes in intermeal interval, rats exhibit limited interindividual variability along this axis, unlike the considerable interindividual variation in ingestive behaviour (Figure~\ref{fig:fingerprint}A). The within-group variation in food intake at similar values of $L_{2}$ is likely to be due to a combination of the natural variability of food intake even when parameters are held constant (see Figure~\ref{fig:termination}C), whereas the between-group variation is due to differences in other behavioural parameters.

\subsection{Ingestive behaviour has a single axis of variation that is modulated by anorectic drugs, photoperiod, and nutritional status}
In order to consume more food in a given bout, an animal can either eat for longer, or it can eat faster. In principle, bout duration and feeding rate could trade off exactly to keep the amount consumed in a bout the same for any bout duration, or could be completely uncorrelated. In practice we found that \textit{feeding rate imperfectly compensates for reduced bout duration}: shorter bout durations were associated with decreased food intake (Figure~\ref{fig:fingerprint}B,C) both within and across groups (see Supplementary Material). Analysis of feeding data from \textit{ad libitum} fed mice in the light and dark period showed an identical pattern (see Supplementary Figure 2), indicating that this tradeoff may apply more generally. Although feeding rate increased at shorter bout durations, it did not do so sufficiently to compensate for the decreased time available to feed, leading to reduced food intake. Longer, slower bouts typically occurred during the dark period, but bout duration and feeding rate showed considerable variation between individuals, drug treatments and between fasted and \textit{ad libitum} fed rats (Figure~\ref{fig:fingerprint}B,C). In order to understand the typical effects of anorectic agents on ingestive behaviour we looked at the group-level posterior probabilities for bout time and feeding rate (Figure~\ref{fig:fingerprint}D-G). These show a probability distribution over the possible values of characteristic feeding rates and bout durations for each condition. We found that PYY\textsubscript{3-36}, LiCl and GLP-1 had a robust effect on feeding rate and bout duration at all doses and in both the light and dark period. \textit{Ad libitum} fed rats have very distinct ingestive behaviours in the dark and light periods, corresponding to long, slow feeding bouts and short, fast bouts, respectively. In a pattern that was found to recur again in this analysis, the effects of anorectic agents was to reduce the difference between these behaviour modes: the day/night difference in parameters of rats given anorectic agents) is less than that in \textit{ad libitum} fed rats. The effect of nutritional status was the same in both photoperiods: bout duration was increased and feeding rate decreased in \textit{ad libitum} fed rats compared to fasted rats during both the dark and light periods. Only LiCl showed a dose-dependent effect on bout duration and feeding rate across the concentrations administered. Of the anorectic drugs administered, GLP-1 had the most pronounced effect on feeding rate, whereas leptin had no effect on either bout duration or feeding rate. This demonstrates that although some of the anorectic agents we investigated do affect ingestive behaviour, their effect differs from that of true satiety induced by \textit{ad libitum} feeding.

\subsection{Fasting and photoperiod strongly affect intermeal interval by altering the effect of stomach fullness, unlike most anorectic agents}
Next, we examined the determinants of intermeal interval duration. Our model allows for both stomach-independent and stomach-dependent control of intermeal interval duration by parameters $L_{1}$ and $L_{2}$ respectively, and the balance of control between these parameters varied strongly in different conditions (Figure~\ref{fig:intermeal interval}A). In the dark period, mean intermeal interval was much shorter, and stomach fullness exerted less control over intermeal interval duration, although it remained the most important factor. The effects of nutritional status and photoperiod were pronounced, with \textit{ad libitum} fed rats in the light period having both longer intermeal intervals and much greater dependence on stomach fullness than \textit{ad libitum} fed rats in the dark period or fasted rats in either photoperiod. PYY\textsubscript{3-36} caused a decrease in $L_{1}$ in both fasted rats in the light period and \textit{ad libitum} fed rats in the dark period, but showed an increase in $L_{2}$ for \textit{ad libitum} fed rats in the dark period only. These changes, although detectable, were small compared to the effects of nutritional status, as indicated by differences between fasted and \textit{ad libitum} fed rats in both photoperiods. GLP-1, LiCl and leptin showed minimal effects on intermeal interval at all doses, with \textit{ad libitum} fed rats given either saline or one of these drugs having almost identical intermeal interval parameters (Figure~\ref{fig:intermeal interval}A). This indicates that although PYY\textsubscript{3-36} may exert anorectic effects by modulating interoceptive sensing \citep{batterham2002gut}, this may not be the means by which other drugs function when administered peripherally (although GLP-1 is known to have satiating properties when administered centrally \citep{hayes2009endogenous}).

\subsection{Complete stomach emptying is the signal for meal initiation in well-fed rats in the light period}
\begin{figure}[htp]
    \centering
    \includegraphics[width=0.95\textwidth]{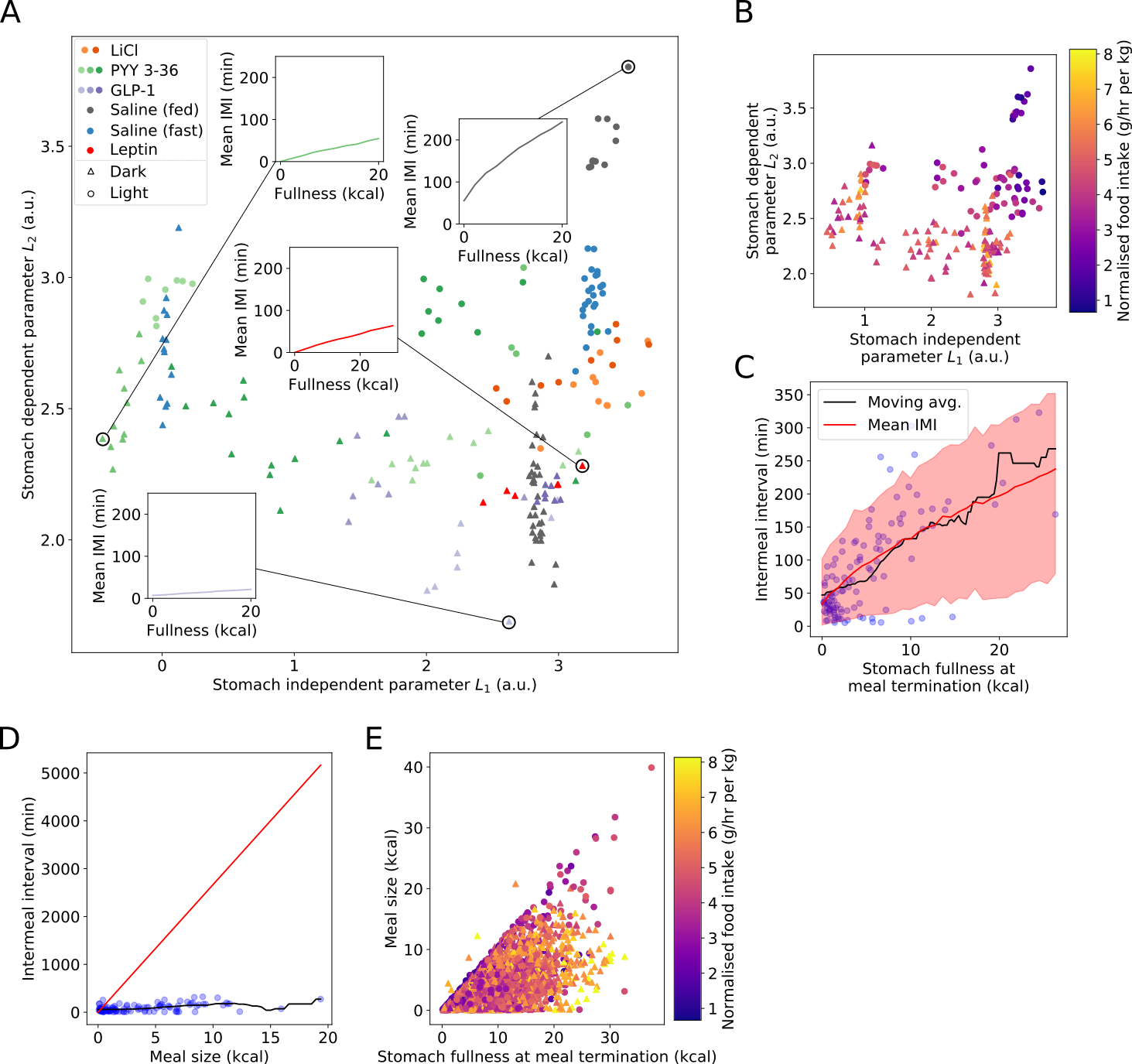}
    \caption{Stomach fullness is predictive of intermeal interval, however the relationship between fullness and intermeal interval varies with photoperiod, fasting status, and anorectic drug administration. The satiety ratio fails to capture the effect of feeding on intermeal interval as it neglects feeding prior to the most recent meal. \textbf{(A)} Intermeal interval parameters are affected strongly by refeeding status, photoperiod, and anorectic drug administration. Intermeal intervals of rats in the light period are strongly affected by stomach fullness, whereas rats in the dark period have briefer intermeal intervals that are less strongly affected by stomach fullness. Rats given lithium chloride behave similarly in the light and dark period. Rats given PYY\textsubscript{3-36} behave more like fasted rats, whereas rats given GLP-1 and leptin have intermeal interval parameters similar to those of \textit{ad libitum} fed rats. Inset figures show how parameter variation affects intermeal interval. \textbf{(B)} Intermeal interval parameters are strongly associated with variation in food intake: animals with lower parameters eat more. Points coloured by normalised food intake. \textbf{(C)} Mean (solid red line) and 95\% posterior predictive interval for \textit{ad libitum} fed rats given saline in the light period. Intermeal interval tracks stomach emptying. Although stomach fullness is indicative of intermeal interval, the predictive window is relatively wide. Blue circles are group data, black line is moving window average. \textbf{(D)} Satiety ratio-based predictions of intermeal interval from meal size are inaccurate. Solid red line shows satiety ratio prediction, blue circles are group data, black line is moving window average. \textbf{(E)} Stomach fullness at meal termination is poorly correlated with meal size, leading to poor predictive ability.}
    \label{fig:intermeal interval}
\end{figure}

Comparing intermeal interval parameters to food intake (Figure~\ref{fig:intermeal interval}B) showed that increases in food intake can be due to decreasing either the stomach-dependent or stomach-independent parameters. The mean intermeal interval closely tracked the time for the stomach to empty in the light period, a phenomenon not observed in other conditions (see Supplementary Material). This may indicate that the main signal to commence feeding in the light period is the complete emptying of the stomach - even low stomach contents were enough to significantly reduce the probability of feeding. In the dark period, however, meal initiation probability showed a graduated response to stomach emptying. Once again, in \textit{ad libitum} fed rats diurnal and nocturnal feeding behaviours differed dramatically, with minimal feeding in the light period and significant food intake occurring in the dark period. As with the within-bout parameters, fasting and the administration of anorectic drugs reduced the strength of this polarisation, increasing the similarity between the light and dark periods.

\subsection{Predictions of intermeal interval from stomach fullness are more accurate than predictions from the satiety ratio}
A comparison of the predicted mean intermeal interval to observed data for \textit{ad libitum} fed rats in the dark period showed a good agreement of model predictions with data (Figure~\ref{fig:intermeal interval}C). However both the data and the predictions showed a substantial degree of variability around the mean. This indicates that we cannot explain all of the variation in intermeal interval using stomach fullness alone. Nevertheless, using stomach fullness to predict intermeal interval is a substantial advance from the satiety ratio, which quantifies satiety as the ratio between the first meal size and the subsequent intermeal interval (see Supplementary Material). The satiety ratio fails to give good predictions across a wide range of meal sizes (Figure~\ref{fig:intermeal interval}D) for a number of reasons. Firstly, it only uses a small amount of the available data as it discards all meals other than the first. Secondly, after the first meal, stomach fullness is poorly correlated with meal size (Figure~\ref{fig:intermeal interval}E) as prior meals have often not been fully digested at meal onset. Finally, the first meal is longer than subsequent ones as it is begun with an empty stomach, which will lead to a shorter than average intermeal interval as the rat has yet to reach its 'typical' fullness.

\subsection{Aversive agents and agents conventionally viewed as satiating have distinct effects on meal termination decisions}
\begin{figure}[htp]
    \centering
    \includegraphics[width=0.95\textwidth]{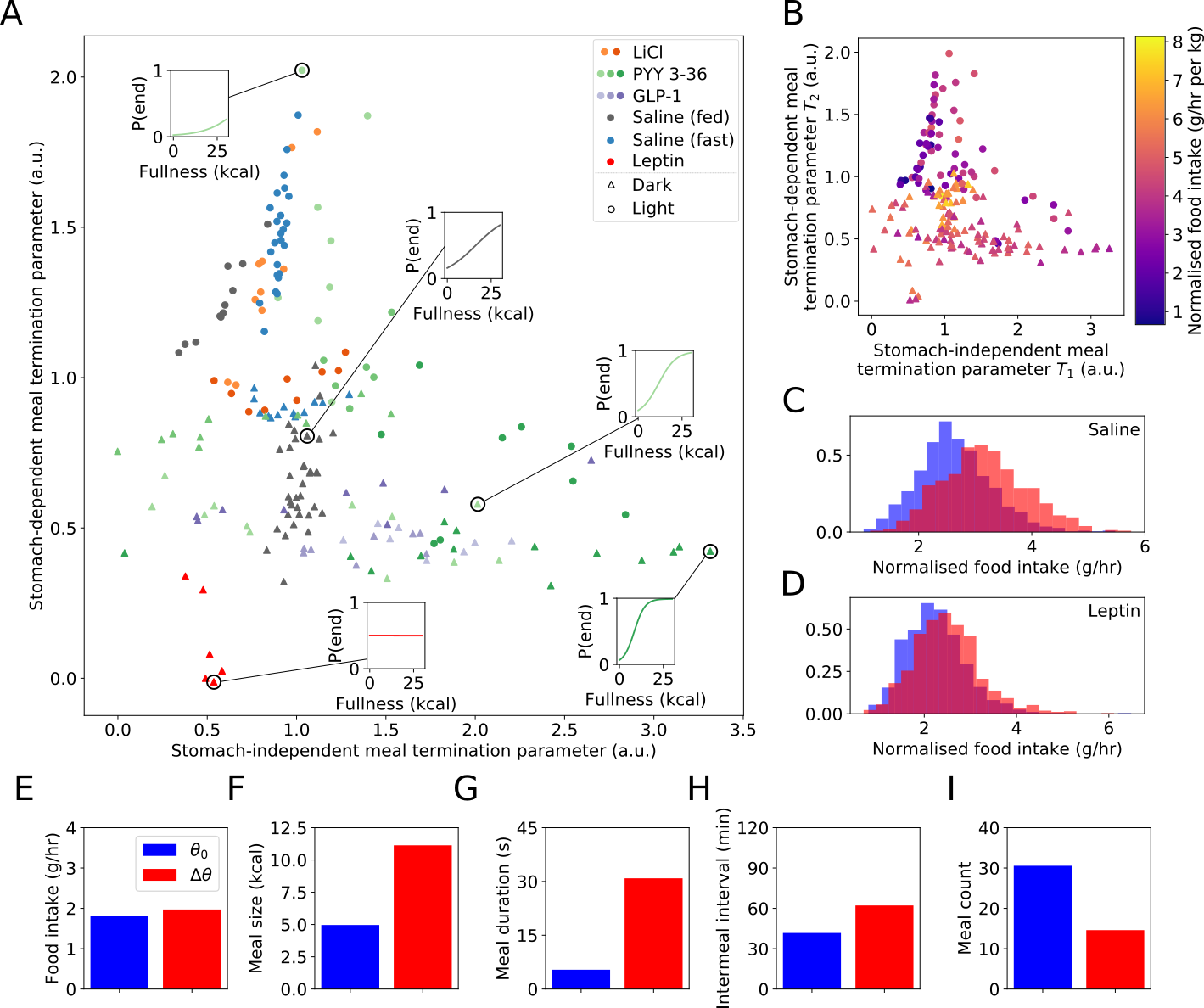}
    \caption{Meal termination decisions vary strongly across experimental conditions, however the effectiveness of altering meal termination decisions to reduce food intake is complex. \textbf{(A)} Aversive agents  (LiCl and high dose PYY\textsubscript{3-36}) lead to meal termination at lower stomach fullness, indicated by a move downwards and to the right. Leptin has a contrasting effect - rats given leptin are more likely to terminate meals at low stomach fullness, and are largely indifferent to stomach fullness in their meal termination decisions. Meal termination is postponed in the dark period, consistent with increased meal size. Individual posterior mean values for stomach-independent and stomach-dependent meal termination parameters $T_{1}$ and $T_{2}$. Insets show posterior mean decision function for representative individuals. \textbf{(B)} Extreme values of either $T_{1}$ or $T_{2}$ are associated with decreased food intake. \textbf{(C, D)} Simulated food intake distributions show that the effectiveness of altering the stomach-dependent meal termination parameter $T_{2}$ varies strongly with the other parameter values (baseline distribution in blue, altered parametrisation in red). Food intake is largely unchanged in simulated rats given leptin when $T_{2}$ is increased, however food intake increases substantially in simulated controls. \textbf{(E-I)} Alterations in multiple parameters simultaneously can produce substantial changes in the microstructure of feeding without affecting overall food intake. Altering mean feeding rate $\mu_{F}$, $T_{1}$ and $T_{2}$ produces an alteration of feeding microstructure similar to that seen in recent studies on CGRP neuron silencing: meals are enlarged and extended \textbf{(F, G)} but intermeal interval is elongated \textbf{(H)} and meal count decreased \textbf{(I)} to fully compensate over 24 hours of feeding.}
    \label{fig:termination}
\end{figure}

Meal termination decisions are believed to be driven at least in part by gastric distention \citep{baird2001parametric}. We modelled the meal termination decision as a sigmoid in stomach fullness, with a prior that allowed for approximately linear, constant, or sigmoid variation across a physiologically-plausible range of stomach fullness. Meal termination decisions varied strongly across the dataset (Figure~\ref{fig:termination}A): in the light period rats given saline had low meal termination probability and weak dependence on stomach fullness, with the other extreme occupied by rats given high-dose PYY\textsubscript{3-36}. Both high-dose PYY\textsubscript{3-36} and LiCl increase meal termination probability and shift termination to occurring at lower stomach fullness, consistent with their nauseating effect \citep{limebeer2010inverse,chelikani2006dose,carter2015parabrachial}. High levels of both meal termination parameters $T_{1}$ and $T_{2}$ were associated with decreased food intake (Figure~\ref{fig:termination}B), but for different reasons. High stomach-dependent termination parameters occurred for rats given saline in the light period; although these rats terminate meals later, they had much longer intermeal intervals and so experienced a net decrease in feeding compared to rats in the dark period. On the other hand, rats given PYY\textsubscript{3-36} terminated meals at a lower stomach fullness, consistent with a a nauseating or otherwise aversive effect. This was found is high-dose PYY\textsubscript{3-36} in both light and dark periods, but not at lower doses, suggesting that our method may be able to identify aversive agents from feeding behaviour. On the other hand, rats given leptin appeared almost indifferent to stomach fullness in their meal termination decisions; they had a high probability of meal termination regardless of stomach fullness. This is in accord with leptin reflecting long term adipose tissue stores rather than acute nutritional status.

\subsection{Changes in meal termination can drastically change feeding microstructure without altering intake}
Recent investigations into meal termination decisions have suggested that feeding behaviour is flexible enough to defend a given food intake under substantial changes to the microstructure of feeding \citep{campos2016parabrachial}. Specifically, following inactivation of Calcitonin Gene-Related Peptide (CGRP)-expressing neurons in the parabrachical nucleus, both meal duration and meal size increased dramatically, with intermeal interval increasing in a compensatory fashion. We investigated whether our model was able to explain these observations by altering the meal termination parameters $T_{1}$ and $T_{2}$, which increase the stomach-dependence of the meal termination probability and make the transition sharper respectively (see Figure~\ref{fig:model_intro}H). Beginning with the parameters from \textit{ad libitum} fed rats given saline in the dark period, which we will denote $\theta_{0}$, we altered $T_{1}$ and $T_{2}$ to cause meal termination to occur at a higher typical level of stomach fullness. In order to recover results similar to those observed in mice with inactivated CGRP neurons, it was also necessary to decrease mean feeding rate $\mu_{F}$. This is consistent with increased bout sizes shown in their representative bout data but which were not quantified. We found that under this change in parameters $\theta_{0}\to\Delta\theta$ (see Supplementary Material) we were able to observe results similar to those found in mice whose CGRP neurons were ablated (see Figure~\ref{fig:termination}E-I); although the absolute values of meal size, duration and other observables differ (as would be expected for different animals), it is possible to replicate the direction and approximate magnitude of change. This indicates the importance of understanding homeostatic regulation of food intake even on short timescales - the effect of stomach fullness on intermeal interval duration is crucial to explaining this behavioural change, and why total food intake is not reduced in these animals.
\subsection{\textit{In silico} studies allow testing of novel behavioural interventions}
In addition to exploring the effects of parameter modifications, using a generative model allowed us to test the effects of behavioural changes as well by modifying the model used to simulate behaviour. We tested the effects of introducing a `refractory period' into the intermeal interval, preventing simulated rats from eating again before a length of time has elapsed (Figure~\ref{fig:controls}A). In this period, meal initiation is prevented, which sets a minimum intermeal interval independent of stomach fullness. The effectiveness of the refractory period depends on the distribution of intermeal interval durations: if the majority of intermeal intervals are longer than the refractory period, then most meal initiations will not be delayed and food intake will will not be substantially reduced. If the refractory period is longer than a substantial fraction of possible intermeal intervals, however, then the average intermeal interval will be extended and food intake will typically be reduced. We found that introducing a 45 minute refractory period was as effective at reducing the food intake of simulated \textit{ad libitum} fed rats in the dark period as high-dose PYY 3-36 (Figure~\ref{fig:controls}B). We also found that the effects of combining the refractory period and PYY\textsubscript{3-36} were additive: introducing a refractory period into the feeding of simulated rats given high dose PYY\textsubscript{3-36} further reduced their feeding, with the difference between \textit{ad libitum} fed rats and those given PYY\textsubscript{3-36} remaining constant as the refactory period was lengthened (Figure~\ref{fig:controls}B).
\subsection{Optimising drug administration protocols can reduce simulated food intake}
\begin{figure}[htp]
    \centering
    \includegraphics[width=0.95\textwidth]{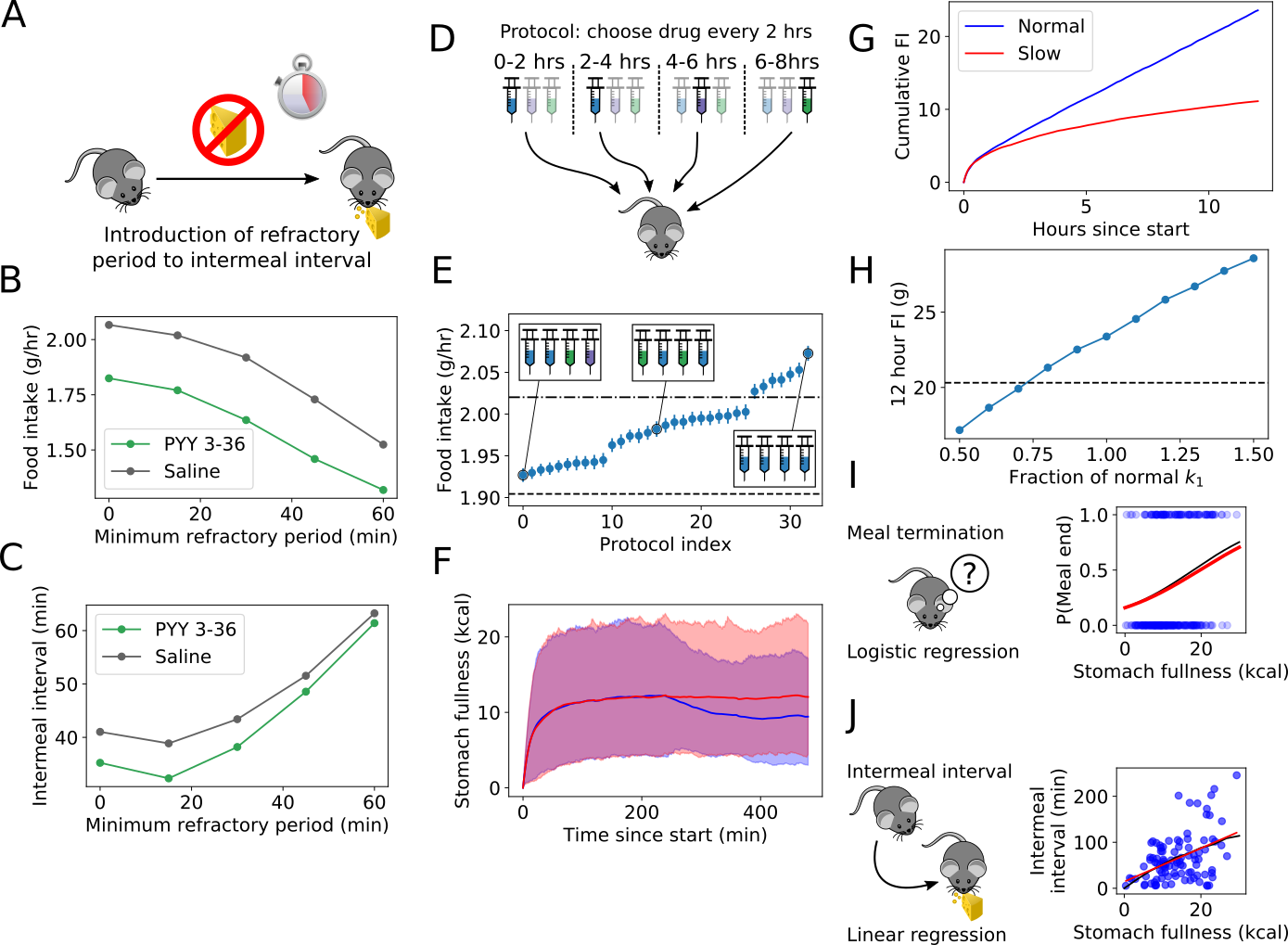}
    \caption{Model-based \textit{in silico} experimentation allows investigation of the effects of parameter changes, design of optimal dosing protocols, and testing of behavioural interventions. \textbf{(A)} Schematic of refractory period experiment: following meal termination, access to food is prevented for a length of time (the refractory period). This is modelled by enforcing the intermeal interval to be at least the refractory period in length. \textbf{(B)} Introducing a refractory period into feeding behaviour reduces food intake to a similar degree as the administration of a high dose of PYY\textsubscript{3-36} in simulated \textit{ad libitum} fed rats in the dark period. \textbf{(C)} Food intake reduction occurs before the mean intermeal interval is substantially reduced, with a surprising dip in the intermeal interval when a short refractory period is introduced. \textbf{(D)} Schematic of optimal dosing experiment. All permutations of drug administrations including at least 2 doses of saline were tested with 10,000 \textit{in silico} repeats. \textbf{(E)} Optimising drug administration schedules can reduce food intake by an additional 7\% when feeding with an initially empty stomach. \textit{Ad libitum fed} high-dose GLP-1, PYY\textsubscript{3-36} and saline parameter values were used to simulate feeding behaviour with drug administration at different times. Error bars indicate standard error of the mean with 10,000 samples. Horizontal lines compare to refractory period reductions in food intake: dash-dotted and dashed lines indicate 15 minute and 30 minute refractory period food intakes for rats given saline respectively. \textbf{(F)} The main effect of the optimal schedule is to reduce food intake once refed, in contrast to optimal dosing during the light period (see Supplementary Materials). Blue and red lines correspond to simulations using optimal administration protocol and control parameters respectively. Shaded area indicates 95th percentile window of stomach fullness. \textbf{(G)} Altering gastric emptying parameter $k$ powerfully reduces food intake once refeeding is complete. A 1/3 reduction in $k$ reduced 12 hour food intake by over 10 grams in simulated \textit{ad libitum} fed rats given saline in the dark period. \textbf{(H)} Altering $k$ over a plausible range linearly reduces food intake in simulated \textit{ad libitum} fed rats given saline in the dark period. Dashed horizontal line indicates food intake typical of rats given high-dose PYY\textsubscript{3-36}. \textbf{(I, J)} Simplifying the model to predict meal termination with a sigmoid and intermeal interval with a linear regression (red lines) against stomach fullness at meal termination shows good agreement with model-derived predictions (black lines).}
    \label{fig:controls}
\end{figure}

We next investigated whether a reduction in food intake could be achieved by optimising the timing of administration of anorectic drugs. Drug administration was simulated by switching to the model parametrisation corresponding to that drug for a 2 hour window. Protocols were constrained to include at least 2 doses of saline in an 8 hour period as we wanted to find the most efficient strategy in order to minimise drug administration (Figure~\ref{fig:controls}D). The candidate anorectic agents were high-dose PYY\textsubscript{3-36} and GLP-1, and parametrisations corresponding to \textit{ad libitum} fed rats in the dark period were used. Unsurprisingly, the least effective strategy was to give no anorectic drugs at all. The most effective strategy consisted of administering anorectic agents at the end of the protocol, giving PYY\textsubscript{3-36} at the 4-hour timepoint followed by GLP-1 at the 6-hour timepoint, although variation in food intake between the ten best protocols was modest (Figure~\ref{fig:controls}E). Following the optimal administration protocol led to a reduction in mean stomach fullness after drug administration until the end of the simulated experiment at the 8 hour timepoint. (Figure~\ref{fig:controls}F).

\subsection{A small reduction in gastric emptying rate efficiently reduces food intake}
Given the degree to which stomach fullness increased intermeal interval across a wide range of conditions, it is plausible that altering gastric emptying rate (e.g. by altering food composition) could be a powerful axis for food intake reduction. Our model allows us to investigate this by altering stomach emptying rate $k$ while holding behavioural parameters $\theta$ constant. We investigated the effects of altering stomach emptying rate in \textit{ad libitum} fed rats in the dark period by using the group posterior mean values of behavioural parameters. Reducing $k$ to $2/3$ of its normal value had a strong, long-lasting effect on 12 hour food intake, but showed very little effect in the first few hours of feeding when beginning at zero stomach fullness (Figure~\ref{fig:controls}G). We then profiled the effect of changing $k$ on 12 hour food intake by varying $k$ between 0.5 and 1.5 times its default value. 12 hour food intake showed an approximately linear dependence on $k$, with a 20\% reduction in gastric emptying showing food intake reduction comparable to high dose PYY\textsubscript{3-36} (Figure~\ref{fig:controls}H). The effects of diet composition on gastric emptying has not been fully characterised, however changes in gastric emptying of this magnitude due to changes in diet have been reported in rats \citep{li2011slower}, mice \citep{whited2004non}, and humans \citep{ma2009effects}.

\subsection{A simplified version of the model can serve as an easy-to-use assay for feeding behaviour}
We investigated possible simplifications of the model in order to make these tools accessible to the largest audience. The aim of this exercise was to find methods that provide most of the insight of the full model we have used in this paper, without requiring Monte Carlo model fits. We began by discarding inter-individual variation by pooling data for all animals within a group, before looking for simplifications of the model. A logistic regression between stomach fullness at bout termination and meal termination probability captures the typical individual's behaviour well (Figure~\ref{fig:controls}G). Because of our previous observation that the intermeal interval depends approximately linearly on stomach fullness in most conditions (although nonlinear fits are possible in the model) we replaced the complex intermeal interval distribution (see Figure~\ref{fig:model_intro}I) with a linear regression between intermeal interval and stomach fullness at meal termination, which accurately reproduced the typical individual in the group (Figure~\ref{fig:controls}J). Once data is pooled across individuals in a group, determining the within-bout parameters reduces to fitting exponential and normal distributions (see Figure~\ref{fig:model_intro}). Using this information it is possible to recover information on each of the components of feeding we have investigated in this paper in most cases, but at the expense of losing individual variation within groups. We have made code to perform these analyses available at \url{https://github.com/tomMcGrath/feeding_analysis}.

\section{Discussion}
We have created a model of feeding to provide a fine-grained understanding of changes in patterns of feeding behaviour, and used this to reveal changes in behaviour between different photoperiods and nutritional states, and under the administration of anorectic drugs. Feeding behaviour within a bout has two possible types of variation: altering bout duration and changing feeding rate. We found that there is a single axis of variation, where feeding rate increases as bout duration decreases, but the increase in feeding rate is only partially compensatory. We found that the anorectic agents GLP-1, PYY\textsubscript{3-36} and lithium chloride mostly act on feeding behaviour within a bout and on meal termination, whereas nutritional status and photoperiod strongly affect intermeal interval as well. Leptin had no effect on bout duration or feeding rate, but had a dramatic effect on meal termination consonant with its effects as a signal of adiposity. Simulation using the model allowed prediction of the effects of changes to behavioural parameters, allowing the recapitulation of recent results on CGRP neuronal ablation \citep{campos2016parabrachial} \textit{in silico}, and predicting that the observed changes in meal termination and intermeal interval are accompanied by changes in within-bout feeding behaviour. Further simulations determined optimal drug administration regimes, and predicted that short-term food intake restriction following meals can be as effective as strong anorectic drugs in reducing food intake, while changes to gastric emptying rate may be even more effective than the anorectic agents studied here.\newline\newline
In recent years a clearer picture of the neuronal circuits underlying feeding behaviour has begun to emerge \citep{sternson2017three}. A simplified description of the current state of knowledge assigns separate behavioural functions to anatomically distinct areas of the brain: AGRP/POMC neurons in the arcuate nucleus are responsible for food-seeking behaviour and meal initiation, but as they are inactivated once food is detected they are not responsible for consummatory behaviour \citep{betley2015neurons, chen2015sensory, mandelblat2015arcuate}. This task is thought to fall to the lateral hypothalamus, which controls consummatory behaviour and reward value of food \citep{jennings2015visualizing, kenny2011reward, rolls2007sensory}. Meal termination decisions involve CGRP-expressing neurons in the parabrachial nucleus \citep{campos2016parabrachial} which is known to incorporate quantitative information on stomach fullness \citep{baird2001parametric}, as we outline below. This experimental evidence is entirely consonant with both the structure of our model, and our findings on the effects of these anorectic drugs on feeding, in that we found that the three separate components of feeding (within-bout behaviour, meal termination, and meal initiation) were affected differently by different anorectic drugs.

\subsection{GLP-1 and Lithium chloride act to alter feeding behaviour within individual bouts}
We found that different drugs had markedly different effects on feeding within bouts: GLP-1 and LiCl had pronounced effects on both bout duration and feeding rate, whereas the effect of PYY and leptin on these parameters was marginal at best. Interestingly, the effect of anorectic agents did not mimic the effects of satiety in the light period. Instead of changing the values of bout duration and feeding rate towards those observed in \textit{ad libitum} fed animals, they instead moved them towards the same values observed in animals given these anorectic agents in the dark period. The difference between true satiety from \textit{ad libitum} feeding and changes in feeding behaviour due to GLP-1 and LiCl was consistent across both photoperiods: a reduction in bout duration and a counterintuitive increase in feeding rate. This suggests that the effects of the anoretic agents investigated is not to mimic satiated behaviour, but to instead \textit{create a third behavioural state that does not represent those found in natural behaviour}. The fact that bout duration and feeding rate are strongly correlated suggests that they are typically regulated together, as if they could be adjusted independently then we would not expect to see the partially-compensatory behaviour observed in Figure~\ref{fig:fingerprint}B.

\subsection{Anorectic agents have weak effects on meal initiation compared to the effect of nutritional status}
Animals with elongated intermeal intervals typically had strongly reduced food intakes. Again, nocturnal and diurnal behaviours in \textit{ad libitum} fed animals represented two behavioural poles, with much longer intermeal intervals occurring in the daytime. Daytime intermeal interval duration was mostly determined by the time taken for the stomach to empty, indicating that the natural organising principle for feeding in the daytime is to wait for complete stomach emptying before feeding recommences. This does not hold true in the dark period, where intermeal intervals are much shorter. Although stomach fullness did increase intermeal interval in the dark period, indicating that interoceptive cues were affecting feeding, the effect is much less pronounced. The anorectic drugs tested appeared to have a less pronounced effect on the intermeal interval than they did on feeding behaviour within the bout: PYY\textsubscript{3-36} at moderate and high doses acted to reduce the stomach-independent component of the intermeal interval, rendering it similar in duration to intermeal intervals observed in fasted rats. On the other hand, GLP-1, LiCl, and leptin had no notable effect on these parameters (Figure~\ref{fig:intermeal interval}A). It was initially surprising that leptin failed to substantially alter the intermeal interval (although it does affect meal termination - see below), however this is likely to be because leptin was administered only to already well-fed rats, and the absence of leptin is much more effective at modulating food intake than the converse \citep{andermann2017toward}. These findings clarify recent results on role of GLP-1 on satiety \citep{gaykema2017activation}: it does reduce food intake, but by altering feeding within meals rather than the length of the intervals between them. Given that there are many extrinsic factors governing meal initiation \citep{strubbe2004timing, petrovich2013forebrain} it is perhaps unsurprising that the intermeal interval is less tightly controlled than other factors, and thus less amenable to change using anorectic drugs. This may make it more easily altered by behavioural interventions, however, which we have found \textit{in silico} can be as effective at reducing food intake as high doses of existing anorectic drugs. On the other hand, agents such as CCK have been shown to increase the intermeal interval \citep{hsiao1979cholecystokinin}, and it remains to be seen whether this is achieved by reducing the basic, stomach-independent drive to feed, or by affecting interoceptive factors.

\subsection{Leptin, and high doses of LiCl and PYY\textsubscript{3-36}, have strong but opposite effects on meal termination}
We observed a wide range of meal termination behaviours (Figure~\ref{fig:termination}A). Rats in the light period typically were less likely to terminate meals, and were less sensitive to stomach fullness, whereas in the dark period meal termination became much more likely, and more sensitive to stomach fullness. Unlike other behavioural parameters, nutritional status had minimal effect on meal termination, whereas anorectic agents had strong effects. High dose lithium chloride increased meal termination probability and rendered rats more sensitive to stomach fullness, as did high-dose PYY\textsubscript{3-36}. Low doses had much less pronounced effects, however. This is consistent with previous observations that high doses of both LiCl and PYY\textsubscript{3-36} can be nauseating \citep{chelikani2006dose,limebeer2010inverse}, and so may make rats less inclined to eat large meals. Leptin's effects were distinct from both short-term nutritional status and the effects of other anorectic agents, and raised the probability of meal termination while decreasing its dependence on stomach fullness. This is in line with leptin's action as a signal of adiposity. The effects of GLP-1 on meal termination were much less pronounced, consistent with the observation that peripheral GLP-1 signalling is not related to gastric distension \citep{hayes2009endogenous}.

\subsection{Recent neuronal ablation studies can be replicated \textit{in silico}}
The structure of food intake is remarkably flexible: it can change dramatically while keeping hourly intake the same - an observation which has recently been made in the context of CGRP ablation studies \citep{campos2016parabrachial}. We investigated whether our model could replicate this phenomenon \textit{in silico} by altering meal termination parameters $T_{1}$ and $T_{2}$. By altering behavioural parameters we were able to create substantial changes in meal size and duration that left total food intake effectively unchanged (Figure~\ref{fig:termination}E-I). In order to create these changes it was also necessary to alter mean feeding rate $\mu_{F}$ - a factor which was not quantified in prior studies. This behavioural plasticity occurs because both meal termination probability and intermeal interval duration vary approximately linearly with stomach fullness in most conditions (Figures~\ref{fig:intermeal interval} and ~\ref{fig:termination}). This linear variation ensures that increases in meal size are compensated for by increases in intermeal duration; increasing typical meal termination fullness by 20\% leads to a 20\% longer typical intermeal interval. This compensatory behaviour defends a constant food intake without needing to finely tune both intermeal interval and meal termination parameters. However, it may be possible to reduce food intake by breaking this linear relationship. This motivated an investigation of the effectiveness of introducing a brief postmeal 'refractory period' in which feeding is forbidden - an intervention that reduced food intake as powerfully as high doses of PYY\textsubscript{3-36} (see below).
\subsection{Behavioural interventions targeting meal initiation may be effective}
We were also able to use our generative model to perform \textit{in silico} experimentation, investigating both the promise of optimising anorectic drug administration and behavioural interventions. Although it was possible to improve food intake reduction by optimising drug administration, the gains we found were relatively small. A simple behavioural intervention, on the other hand, was highly effective at reducing feeding. By introducing a `refractory period' of 45 minutes during which meal initiation was forbidden we were able to achieve food intake reduction approximately equivalent to that evoked by a high dose of PYY\textsubscript{3-36} in the same conditions (Figure~\ref{fig:controls}B).
\subsection{Slowing gastric emptying potently reduces food intake}
By changing the stomach emptying parameter $k$ in our simulation while keeping behavioural parameters fixed, we were able to investigate the effect of slowing gastric emptying, as might be caused by altering diet composition. Physiologically plausible changes in the gastric emptying parameter $k$ led to food intake reduction of the same or greater magnitude than that following the administration of anorectic drugs (Figure~\ref{fig:controls}G,H). Because changes in gastric emptying primarily affect the intermeal interval duration, whereas we have shown that the anorectic drugs investigated here have their strongest effects on ingestive behaviour and meal termination, it is plausible that changes in gastric emptying could work in concert with anorectic agents to have an increased effect on food intake reduction. Both GLP-1 and PYY\textsubscript{3-36} are believed to decrease gastric emptying rate acutely \citep{wettergren1993truncated, batterham2002gut} although our results demonstrate that this is not their only effect (and may not be their main effect long-term), as they also strongly alter feeding rate and bout duration. Our model cannot currently detect changes in gastric emptying rate directly from bout-level data, however these \textit{in silico} experiments indicate that delaying gastric emptying is a powerful way to reduce total food intake.
\subsection{Summary}
We have constructed a stochastic model of feeding that accurately recapitulates observed behaviour. We validated this model and used it on a substantial dataset of rat feeding in a wide variety of conditions, including across both photoperiods, both \textit{ad libitum} fed and fasted, and with the administration of a variety of anorectic agents. We also demonstrated the applicability of this model to mouse feeding data with only minimal parameter changes. Using our model we obtained deeper, more finely-grained insights into the diverse behavioural effects of anorectic agents, and how they reduce (or fail to reduce) food intake in different conditions. Understanding the diverse behavioural effects of anorectic agents demonstrated that they typically do not mimic `natural' satiety induced by \textit{ad libitum} access to food. We found that the precise behavioural effects of the anorectic agents we studied were as follows:
\begin{itemize}
\item GLP-1: reduce bout duration and increased feeding rate in \textit{ad libitum} fed rats at night, no effect on meal termination or intermeal interval parameters.
\item PYY\textsubscript{3-36}: diverse effects in all aspects of feeding. PYY\textsubscript{3-36} reduced bout duration and increase feeding rate in \textit{ad libitum} fed rats at night, with the opposite effect in fasted rats during the day (suggesting an alternative behavioural state not representative of typical behaviour). Meal termination occurred at lower stomach fullness in both light and dark period and intermeal interval increased.
\item Leptin: strong alterations in meal termination. Meal termination probability increases, especially at low stomach fullness, no effects on other aspects of feeding.
\item LiCl: Decrease in feeding rate and dependence of intermeal interval on stomach fullness. Meal termination shifted to lower stomach fullness.
\end{itemize}
We found what appears to be a universal factor in feeding behaviour: as the duration of a bout shortens, the rate of feeding within the bout increases. This tradeoff does not perfectly maintain food intake, however, as the increase in feeding rate is insufficient to maintain the amount eaten within the bout. This observation held true across all conditions, and was also observed in mice. The second reliable determinant of food intake was the factor determining how stomach fullness affects the duration of the intermeal interval, reinforcing the necessity of using a model that tracks stomach fullness over time. Using a dynamic calculation of stomach fullness, rather than simply tracking the size of individual meals, led to a dramatic improvement in predictive power over the satiety ratio.\newline\newline
Finally, we used our computational model to perform \textit{in silico} experiments investigating the effects of other interventions on food intake. We examined optimal drug adminstration protocols, the introduction of a postmeal refractory period, and altering gastric emptying rate. Although drug administration protocols can be optimised \textit{in silico}, the effect is much weaker than either introducing a refractory period or altering gastric emptying. A 45 minute refractory period or a 20\% decrease in gastric emptying rate were both sufficient to reduce 12 hour food intake as much as a high dose of PYY\textsubscript{3-36}, highlighting the potential of alternative interventions to reduce food intake.\newline\newline
We have made the code used in the analysis freely available online and have also released simplified analytical tools for easy analysis of bout feeding data. Although we believe this model is a powerful tool for gaining a finer-grained understanding of feeding behaviour, more work remains to be done to enhance the model. For instance, incorporating longer term variations in behaviour due to energy balance and energy expenditure, through a more complex model of the gut \citep{dalla2007meal, janssen2011role} or by incorporating effects of anorectic agents on gastric emptying rate (one way by which both GLP-1 and PYY\textsubscript{3-36} are believed to reduce food intake). The analysis we hve done could also be extended to incorporate other factors we have held constant here, for instance food palatability or environmental temperature. Another avenue of research is to use rich data such as videos in conjunction with bout-level feeding data to connect feeding behaviour to other aspects of behaviour, for instance by recognising nausea from behavioural signals \citep{andrews2006signals}. This could be accomplished by combining graphical models (such as the model we have used in this paper) with neural networks, an approach that has shown promise in behavioural analysis \citep{johnson2016composing}. This promising avenue of research could make use of the increasing range of high-resolution data becoming available \citep{sternson2016emerging,dhawale2017automated}. Potential applications for our model could be to characterise the behavioural effects of opto- or chemogenetic changes in brain areas responsible for the regulation of food intake, to guide searches for synergistic combinations of anorectic agents, and to understand the effects of multiple treatments in combination.
\section{Methods and Materials}
\subsection{Piecewise Deterministic Markov Processes}\label{ss:PDMP}
Piecewise deterministic Markov processes (PDMPs), also known as stochastic hybrid models, are a generalisation of Markov chains that capture variation in transition rates between states due to some continuously-varying parameter or set of parameters \citep{davis1984piecewise}. The evolution of these parameters are, in turn, determined by the state of the system. In our model, the discrete states are the behavioural states (feeding $F$, short within-meal pause $S$ and long intermeal interval $L$), and the continuously-varying parameter is stomach fullness $x(t)$. A final distinction between PDMPs and Markov chains is that the transition rate quantifies transitions \textit{out of} a state, rather than into a new state.
\subsubsection{Stomach fullness equation}
Stomach fullness $x$ is deterministic when conditioned on the behavioural state $s \in \{F, S, L\}$, and is given by (see Figure~\ref{fig:model_intro}):
\begin{equation}
\dot{x}(x, t, s) = \begin{cases}
               \rho & \textrm{s = F}\\
               -k\sqrt{x} & \textrm{s = S, L}
            \end{cases}
\label{eq:stomach}\end{equation}
where $\rho$ is randomly sampled for each feeding bout (see below and Figure~\ref{fig:model_intro}D). The parameter $k = 0.0055$ was fitted for male Wistar rats when this stomach emptying model was first defined \citep{booth1978hunger}. Equation~\ref{eq:stomach} neglects digestion during feeding to allow for a simpler mathematical formulation that yields analytically-solvable transition rate equations. Feeding bouts are typically much shorter than intermeal pauses (where the vast majority of digestion occurs) so this approximation is well-justified.

\subsubsection{State lifetimes}
The duration of a behavioural state $s$ is a random variable whose distribution is determined by the transition rate out of that state $\lambda_{s}(x, \theta)$, where $\theta$ is a vector of parameters and $x$ is the stomach fullness. If the transition rate out of a state in an infinitesimal interval $[t, t+\delta t]$ is given by $\lambda_{S}(x(t))\delta t + o(\delta t)$ and the process begins in state $x_{0}$ at time $t_{0}$ then the probability density function for state lifetimes is given by
\begin{equation}
    p(t|x_{0}, \mathbf{\theta}) = \lambda_{s}(x(t, x_{0}),\, \mathbf{\theta})e^{-\int^{t}_{0}\lambda_{s}(x(\tau, x_{0}), \,\mathbf{\theta})d\tau}
\end{equation}
Setting $\lambda$ to be constant yields the exponential distribution. We define the rates as follows (see Figure~\ref{fig:model_intro}):
\begin{align*}
    \lambda_{F}(x) &= \lambda_{F}\\
    \lambda_{S}(x) &= \lambda_{S}\\
    \lambda_{L}(x) &= \frac{1}{L_{1} + L_{2}x},
\end{align*}
where the $\lambda_{L}$ equation captures our intuitive understanding that feeding is more likely to recommence when the stomach is empty, but allows for stomach independence in the $L_{2}\to 0$ limit, in which case the intermeal pauses will also become exponential.
\subsubsection{Transition kernel}
After a state finishes (for instance the end of a bout) it is necessary to determine what happens next. The transition kernel is a parametrised function that gives the probability of transitioning from state $s$ to new state $s'$ (for example, the probability of transitioning from feeding into the within-meal pause state). The transition probabilities are:
\begin{equation}
p(s'|s,x, T_{1}, T_{2}) = \begin{cases}
               1 & s=S, s'=F\\
               1 & s=L, s'=F\\
               \frac{1}{1 + e^{-T_{1}(x - T_{2})}} & s=F, s'=L\\
               1-\frac{1}{1 + e^{-T_{1}(x - T_{2})}} & s=F, s'=S\\
               0 & \textrm{otherwise.}
            \end{cases}
\end{equation}
For clarity, deterministic transitions are not shown in Figure~\ref{fig:model_intro}, only the stochastic transition between feeding and the pause states $S$ and $F$ are indicated.

\subsection{Bayesian hierarchical model}
\subsubsection{Introduction}
In order to maximise our inferential power, as well as distinguish between interindividual and group variation, we use a Bayesian hierarchical model. This is a standard technique in Bayesian inference where parameters for a given individual are modelled as being drawn from some distribution applying to the whole group, and we aim to infer both the individual and group parameters. Extensive explanations of this technique are available in a number of books, of which we particularly recommend \citep{gelman2014bayesian, mcelreath2015statistical}, however we will provide a brief introduction here. 
\newline\newline
In a hierarchical model, parameters are separated into two (or more) sets: each group (e.g. \textit{ad libitum} fed rats in the dark period) has a set of hyperparameters $\mathbf{\psi}$, which represent group-level variation, and individuals each have a set of parameters $\mathbf{\theta}$, which are drawn from $\mathbf{\psi}$. Now, by Bayes' theorem, the posterior probability for the complete parameter set $\{\mathbf{\psi},\mathbf{\theta}\}$ conditioned on the data $y$ is given by:
\begin{equation*}
    p(\psi, \theta | y) \propto p(y | \theta)\,p(\theta | \psi)\,p(\psi),
\end{equation*}
so the group level variation depends on the data through the 'lower level' model. More concretely, in this case each individual (indexed here by $j$) has a dataset we denote by $y_{j}$. We model this dataset using a PDMP with a different parameter vector $\theta^{j}$ for each individual, and denote a single parameter $i$ in this parameter vector by $\theta^{j}_{i}$. Group level variation is introduced by modelling $\theta^{j}_{i}$ as being drawn from a multivariate normal distribution, in a way that we will make concrete in the following subsection. 

\subsubsection{Model formulation}
The group level distribution was modelled as a multivariate normal with mean $\mu$ and covariance matrix $\Sigma$ constructed by a separation strategy using a LKJ correlation matrix and independent variance parameters $\tau_{i}$:
\begin{align*}
    \tilde{\theta}^{j} &\sim \mathcal{N}(\mu, \Sigma)\\
    \Sigma &\sim \text{diag}(\tau)\,\Omega(\nu)\,\text{diag}(\tau)\\
    \tau_{i} &\sim \text{HalfCauchy}(2.5)\\
    \mu &\sim \mathcal{N}(\mu^{*}, 2)
\end{align*}
where $\text{diag}(\tau)$ is a diagonal matrix whose elements are drawn from a Half-Cauchy prior as shown above, $\tilde{\theta}^{j}$ is is the vector of parameters for individual $j$: $\tilde{\theta}^{j} = \{\lambda_{F}^{j}, \rho_{F}^{j}, \mu_{F}^{j}, \sigma_{F}^{j}, \lambda_{S}^{j}, T_{1}^{j}, T_{2}^{j}, L_{1}^{j}, L_{2}^{j} \}$. This strategy has been shown to be superior to Inverse-Wishart covariance matrix priors, which introduce biases into estimates of covariance \citep{alvarez2014bayesian}. The mean vector $\mu^{*} = (-3, -3, -3, 1, 1, -1, 3, 3)$ was introduced following inspection of the \textit{ad libitum} fed saline datasets, and was used in conjunction with a diffuse covariance matrix to yield a weakly-informative prior which supported both strong and weak dependence of the intermeal interval duration and meal termination on stomach fullness. With the exception of parameters $T_{1}$ and $T_{2}$ (the transition kernel parameters), all group-level parameters were inferred on a log10 scale.

\subsubsection{Inference}
We used the NUTS auto-tuning Hamiltonian Monte Carlo (HMC) sampler implemented in PyMC3 \citep{salvatier2016probabilistic} to generate posterior samples in order to overcome the posterior curvature issues generically associated with hierarchical Bayesian models. It was also necessary to use the non-centred parametrisation suggested by Betancourt and Girolami \citep{betancourt2015hamiltonian} to overcome these issues. 5,000 tuning samples were generated and discarded to effectively parametrise NUTS, with 5,000 further samples taken and retained once tuning was achieved. It is worth noting that HMC samplers require a much lower sample size than traditional Metropolis-Hastings samplers, due to their more rapid exploration of the posterior. MAP initialisation was not used as the typical set for hierarchical models is often far from the MAP point \citep{betancourt2015hamiltonian}.

\subsection{Experimental procedures}
\subsubsection{Animals}
Male Wistar rats weighing between 254 and 547 grams were individually housed under controlled temperature (21-23$^{\circ}$C) and humidity on a 12h light: 12h darkness cycle, with the light cycle between 0600 and 1800. All animals had ad libitum access to standard chow RM1 (SDS, Witham, UK) and water unless stated otherwise. Rats were acclimatised to all experimental procedures, including intraperitoneal injection. All animal procedures were in accordance with the UK Home Office Animals (Scientific Procedures) Act 1986 and were approved by the Animal Welfare and Ethics Review Board at the Central Biological Services unit at the Hammersmith Campus of Imperial College London.

\subsubsection{Feeding data collection and preprocessing}
Rats were individually placed into a 24-chamber open-circuit comprehensive laboratory animal-monitoring system (CLAMS, Columbus Instruments, Columbus, OH, USA) and acclimatised to the system for 24h prior to fasting. In protocols 1 and 2 (shorter-duration studies involving the administration of anorectic drugs), animals were acclimatised to intraperitoneal injections prior to the experiment. Quantitiation of feeding was carried out using the feeding bout data reported by CLAMS. This was processed prior to analysis in order to remove erroneous data (negative and cancelling readings, low and high outliers). This processing can be reproduced using the data and notebooks available at \url{https://github.com/tomMcGrath/bayesianBehaviour}. The number of animals reported in Table~\ref{tab:drugs} is the number remaining after data cleaning (see Supplementary Materials).

\subsubsection{Protocol 1}
Animals were allowed to feed \textit{ad libitum} both prior to and throughout the experiment. In the early dark period anorectic drugs were administered via intraperitoneal injections at the doses and times in Table~\ref{tab:drugs}.

\subsubsection{Protocol 2}
Following an overnight fast, animals received an intraperitoneal injection of saline or an anorectic drug (see Table~\ref{tab:drugs}) and were subsequently allowed to feed \textit{ad libitum} from the early light period. Injections were carried out according to the schedule below, allowing for the study of feeding behaviour in both photoperiods.

\subsubsection{Protocol 3}
Following an overnight fast, animals were allowed to feed \textit{ad libitum} from the onset of the light period. Data from this protocol was recorded as a continuous 72 hour recording, segmented according to the times in Table~\ref{tab:drugs}. After 24 hours of \textit{ad libitum} feeding following the overnight fast, rats were assumed to be refed, and so were classified as \textit{ad libitum} fed for the remaining two light/dark cycles.

\subsubsection{Drug administration}
Anorectic drugs, doses, and administration schedules are given in Table~\ref{tab:drugs}.
\begin{table}
\begin{footnotesize}
\begin{center}
\begin{tabular}{ c c c c c c c }
\textbf{Drug} & \textbf{Dose} & \textbf{Start time} & \textbf{Duration} & \textbf{Prior feeding} & \textbf{\# animals} & \textbf{Protocol}\\
GLP-1 & 30 nmol/kg & 1900 & 8hrs & \textit{ad libitum} & 8 & 1\\
GLP-1 & 100 nmol/kg & 1900 & 8hrs & \textit{ad libitum} & 8 & 1\\
GLP-1 & 300 nmol/kg & 1900 & 8hrs & \textit{ad libitum} & 9 & 1\\
PYY\textsubscript{3-36} & 1.5 nmol/kg & 1900 & 8hrs & \textit{ad libitum} & 12 & 1\\
PYY\textsubscript{3-36} & 100 nmol/kg & 1900 & 8hrs & \textit{ad libitum} & 11 & 1\\
PYY\textsubscript{3-36} & 300 nmol/kg & 1900 & 8hrs & \textit{ad libitum} & 14 & 1\\
Leptin & 2 mg/kg & 1900 & 8hrs & \textit{ad libitum} & 6 & 1\\
Saline & n/a & 1900 & 8hrs & \textit{ad libitum} & 24 & 1\\
LiCl & 32 mg/kg & 0900 & 8hrs & fasted overnight & 10 & 2\\
LiCl & 64 mg/kg & 0900 & 8hrs & fasted overnight & 9 & 2\\
PYY 3-36 & 1.5 nmol/kg & 0900 & 8hrs & fasted overnight & 8 & 2\\
PYY 3-36 & 100 nmol/kg & 0900 & 8hrs & fasted overnight & 6 & 2\\
PYY 3-36 & 300 nmol/kg & 0900 & 8hrs & fasted overnight & 9 & 2\\
Saline & n/a & 0900 & 8hrs & fasted overnight & 13 & 2\\
Saline & n/a & 0600 & 10hrs & fasted overnight & 11 &  3\\
Saline & n/a & 1800 & 10hrs & fasted overnight & 11 &  3\\
Saline & n/a & 0600 & 10hrs (x2) & \textit{ad libitum} & 11 & 3\\
Saline & n/a & 1800 & 10hrs (x2) & \textit{ad libitum} & 11 & 3\\
\end{tabular}
\end{center}
\end{footnotesize}
\caption{\label{tab:drugs}Details of drug doses and administration protocols. Light on/off times were 0600 and 1800 respectively. Anorectic agents were Glucagon-like peptide 1 (GLP-1), Peptide YY\textsubscript{3-36} (PYY\textsubscript{3-36}), lithium chloride (LiCl) and leptin.}
\end{table}

\section{Acknowledgments}
We thank Juvid Aryaman and Hanne Hoitzing for helpful discussions.

\section{Additional Information}
\subsection{Funding}
TMM is supported by BBSRC studentship BB/J014575/1. ES was funded by an NC3Rs studentship. NSJ is funded by EPSRC grant EP/N014529/1. The Section of Endocrinology and Investigative Medicine is funded by grants from the NC3Rs, MRC, BBSRC, NIHR, Innovate UK (Technology Strategy Board), the Society for Endocrinology, an Integrative Mammalian Biology (IMB) Capacity Building Award and an FP7 HEALTH-2009-241592 EuroCHIP grant, and is supported by the NIHR Biomedical Research Centre Funding Scheme. 

%\nocite{*} % This command displays all refs in the bib file
\bibliographystyle{model1-num-names}
\bibliography{ms.bib}

\newpage
\vspace{0.1\paperheight} {\Large \textbf{Interpreting, forecasting, and controlling feeding behaviour using high-resolution data - Supplementary Material} } \newline 
\\ \textbf{Thomas M McGrath, Eleanor Spreckley, Aina Fernandez Rodriguez, Carlo Viscomi, Amin Alamshah, Elina Akalestou, Kevin G Murphy, Nick S Jones} \setcounter{page}{1} \renewcommand{\theequation}{Equation S\arabic{equation}} \renewcommand{\eqref}[1]{Equation \ref{#1}} \makeatletter \DeclareRobustCommand{\gobblesomeargs}[9]{#9} \renewcommand{\p@equation}{\gobblesomeargs} \makeatother \setcounter{equation}{0} \renewcommand\thefigure{S\arabic{figure}} \setcounter{figure}{0} \renewcommand\thetable{S\arabic{table}} \setcounter{table}{0} \renewcommand\thesection{S\arabic{section}} \setcounter{section}{0}

\section{Data and software availability}
Raw and preproccessed data is available as a Mendeley dataset (\url{doi:10.17632/vpm89vrz7g.1}), and all of the code used to process the data, perform inference and generate figures is available as a Github repository at \url{https://github.com/tomMcGrath/feeding_analysis}.

\section{Data preprocessing}
Bout data was cleaned according to the following criteria:
\begin{itemize}
\item Bout size $< 4$ grams
\item Bout duration $< 1000$ seconds
\item Feeding rate $< 0.02$ grams per second
\item Bout duration $> 4$ seconds.
\end{itemize}
In addition, bouts of negative size and bouts that were immediately cancelled by a subsequent bout were also removed. Data for a rat was removed if any of the following criteria held:
\begin{itemize}
\item More than 25 cancelling bouts
\item More than 200 negative-value bouts
\item Less than 5 bouts
\item More than 30 bouts excluded by the bout cleaning criteria above.
\end{itemize}
This procedure led to removal of 43 out of 198 files from the original dataset.

\section{Model validation via simulation}
The groups used in model design were low, medium and high-dose PYY given to \textit{ad libitum} fed rats in the dark period, and their associated controls. All other data was held out and not used in the model design phase. Priors were obtained via Empirical Bayes from this dataset. Simulations were performed by Monte-Carlo resampling the individual posterior parameter distribution for each rat 100 times, matching initial stomach fullness and behaviour sequence duration. The mean of the simulated food intake distribution was compared against the actual food intake in order to produce Figure 1H.

\section{Reduction of food intake by anorectic agents}
Figure~\ref{fig:FI} shows average normalised food intake for each group. In order to account for differences in bodyweight and experiment duration between experimental groups, food intake was normalised to units of grams per hour per kilogram of bodyweight. Bodyweight values were recorded at the beginning of each experiment. Rats recorded as recovering from a fast in the dark period were subject to an overnight fast before refeeding was permitted in the light period. Although feeding was elevated following the overnight fast (Figure~\ref{fig:FI}), normalised FI was lower than dark period values, so we still consider these rats to be in a nutritional deficit.
\begin{figure}
\centering
\includegraphics[width=0.5\textwidth]{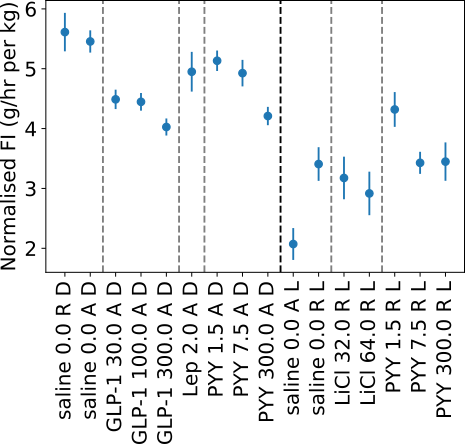}
\caption{Anorectic agents reduce food intake in a dose-dependent manner. Dark dashed line separates light and dark period data, light dashed lines separate different anorectic agents. Error bars show standard error of the mean.}
\end{figure}\label{fig:FI}

\section{Inference for mouse data}
\subsection{Experimental procedures}
Mouse behavioural data was obtained from both male and female wild-type C57BL/6 mice between 17.8 and 36.2 grams in the light and dark periods. Light hours were 0600-1800. Animals were allowed \textit{ad libitum} access to standard chow (R105, Safe Diets, Augy, France) throughout. Mice were acclimatised to all experimental procedures. All procedures were conducted under the UK Animals (Scientific Procedures) Act, 1986, approved by Home Office license (PPL: P6C97520A) and local ethical review. The animals were maintained on a C57BL/6 background at 19-21 C in a temperature- and humidity-controlled animal-care facility, with a 12 hr light/dark cycle and free access to water and food.

\subsection{Data preprocessing}
Data preprocessing was carried out using the same code as the rat data, with the following changes in exclusion criteria due to the increased length of the behaviour sequences:
\begin{itemize}
\item More than 50 cancelling bouts
\item More than 200 negative-value bouts
\item Less than 5 bouts
\item More than 200 bouts excluded by the bout cleaning criteria above.
\end{itemize}

\subsection{Data analysis}
Group level priors were set via Empirical Bayes analysis of fully pooled data. Based on this analysis, group means were set to:
\begin{equation}
\theta = [-2,-4,-4,1,1,-1,3,4]
\end{equation}
To accomodate the decreased size of mouse stomachs, we reparametrised the meal termination function to
\begin{equation}
\textrm{Prob}(end) = (1 + \exp(-0.1T_{1}(x-T_{2}))^{-1}.
\end{equation}
Apart from these change in parameters, inference was carried out using the same code as the rat data (see Materials and Methods), using the NUTS sampler with 5,000 tuning and 5,000 sample steps.

\subsection{Results}
\begin{figure}
\centering
\includegraphics[width=0.5\textwidth]{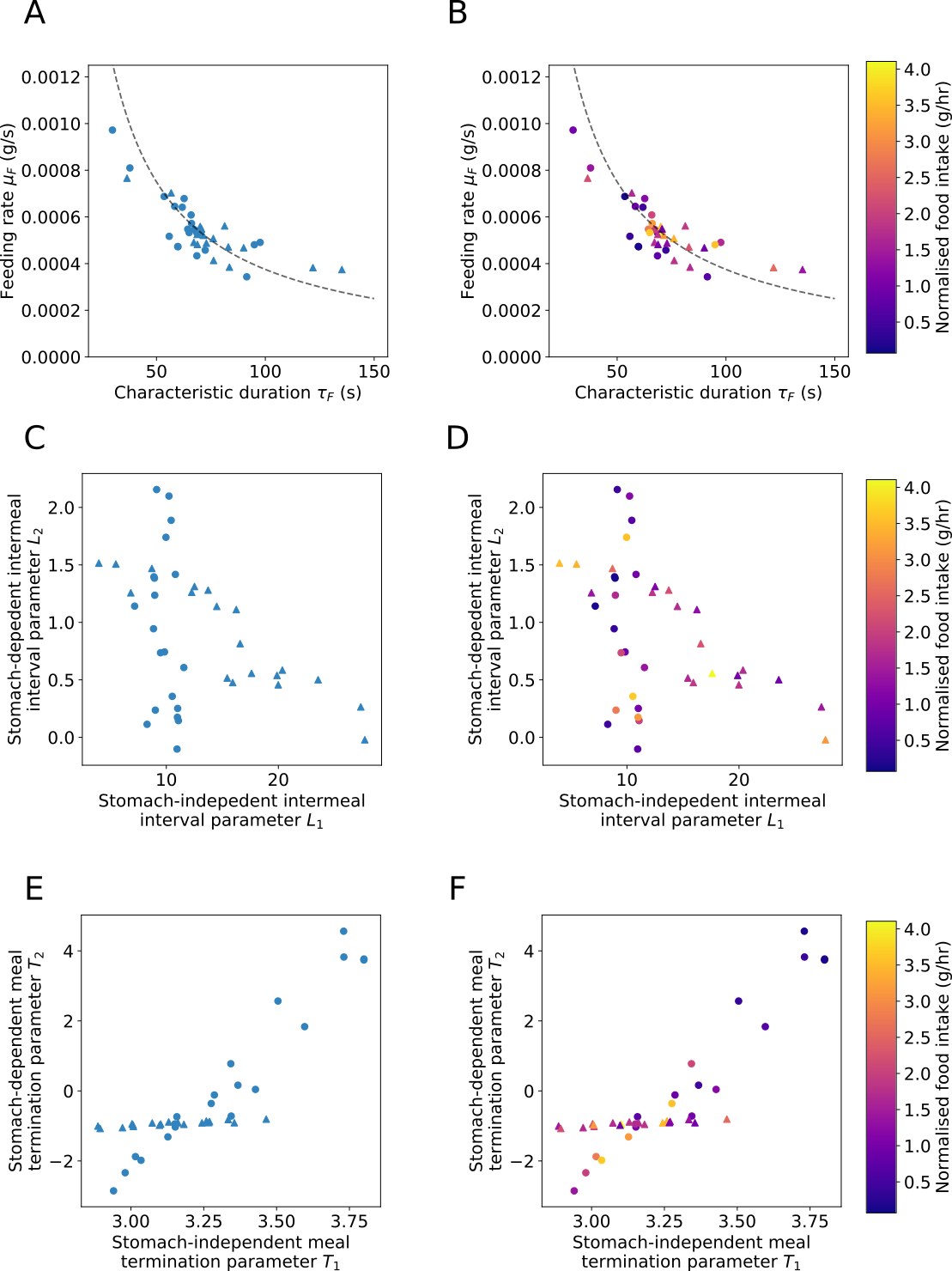}
\caption{The stochastic feeding model can also be applied to mouse data. Right column indicates bodyweight-normalised food intake. \textbf{(A,B)} The imperfect duration/feeding rate tradeoff appears in both the dark and light period for mice. \textbf{(C,D} Mice show a wide degree of variation in $L_{2}$. \textbf{(E,F)} Meal termination parameters are correlated in mice, and low $T_{1}$ and $T_{2}$ are both associated with high food intake.}\label{fig:supp_mice}
\end{figure}

See Figure~\ref{fig:supp_mice}A-F.

\section{The bout duration/feeding rate tradeoff holds within groups}
In order to ensure that the characteristic bout duration/mean feeding rate tradeoff we observed held both within and across groups, we examined feeding data for each anorectic agent separately. With the exception of leptin, which shows limited variation in either parameter, this trend held consistently across different conditions (Figure~\ref{fig:group_tradeoff}).
\begin{figure}[h!]
\centering
\includegraphics[width=0.95\textwidth]{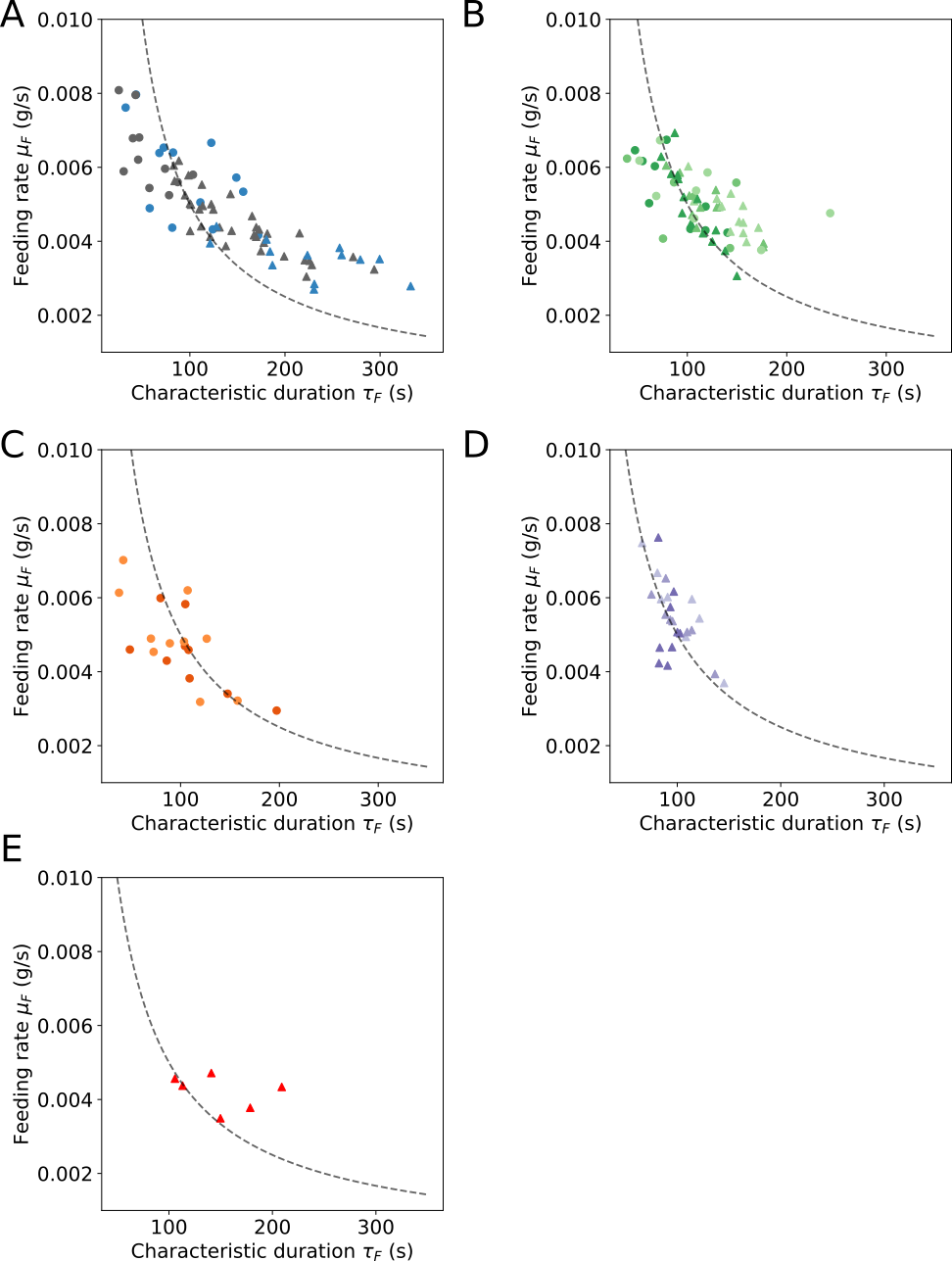}
\caption{Group-level plots for characteristic bout duration and feeding rate. Symbol type indicates photoperiod. Triangles: dark period, circles: light period. \textbf{(A)} Fasted and \textit{ad libitum} fed rats given saline in the dark and light periods. \textbf{(B)} Rats given Peptide YY 3-36 (PYY 3-36) in the light (fasted) and dark (\textit{ad libitum} fed) periods. \textbf{(C)} Lithium Chloride recovering from a fast in the light period. \textbf{(D)} \textit{Ad libitum} fed rats given Glucagon-like peptide 1 (GLP-1) in the dark period. \textbf{(E)} \textit{Ad libitum} fed rats given Leptin in the dark period.}
\end{figure}\label{fig:group_tradeoff}

\section{Calculating the satiety ratio}
The satiety ratio was calculated for each group using the first meal from each animal's data. Meals were determined using a hard cutoff in order to avoid including assumptions from our Bayesian model: bouts followed by pauses of less than 300 seconds were grouped into meals. The group satiety ratio was calculated as the average of the individual satiety ratios - for a group of size $N$, the satiety ratio $r$ is given as the average of the individual satiety ratios $r_{i}$. These are determined from the first meal size $s^{0}_{i}$ and intermeal interval $t^{0}_{i}$
$$
r = \frac{1}{N}\sum_{i=1}^{N}r_{i} = \frac{1}{N}\sum_{i=1}^{N}\frac{s^{0}_{i}}{t^{0}_{i}}.
$$
Intermeal interval predictions $\tilde{t}_{i}$ were generated for subsequent meals of size $s_{i}$ using the formula $\tilde{t}_{i} = rs_{i}$.

\section{Mean intermeal interval tracks stomach emptying time in the light period}
Stomach emptying time $t_{0}$ was calculated using the formula $t_{0} = 2\sqrt{x_{0}}/k_{1}$, where $x_{0}$ is the stomach fullness at the beginning of the intermeal interval and $k_{1}$ is the stomach emptying rate parameter. These were compared to intermeal intervals sampled using the group mean parameters for \textit{ad libitum} fed rats in the light and dark period to generate Figure~\ref{fig:IMI_emptying}. The light period shows nonlinear dependence of intermeal interval on initial stomach fullness in a way that matches stomach emptying time, whereas the dark period shows linear dependence on initial stomach fullness.
\begin{figure}[h!]
\centering
\begin{subfigure}{0.45\textwidth}
\centering
\includegraphics[width=\textwidth]{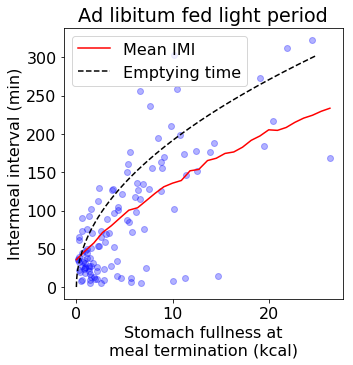}
\end{subfigure}
~
\begin{subfigure}{0.45\textwidth}
\centering
\includegraphics[width=\textwidth]{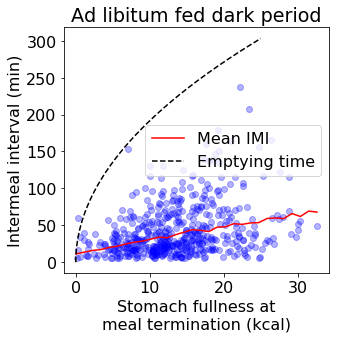}
\end{subfigure}
\caption{Intermeal interval tracks stomach emptying in \textit{ad libitum} fed rats in the light period but not the dark. Blue circles indicate intermeal interval data, red line indicates the mean intermeal interval (obtained from Monte Carlo simulation) and the dashed line shows the time the stomach would take to empty if no feeding took place.}
\end{figure}\label{fig:IMI_emptying}

\section{Replicating CGRP ablation studies \textit{in silico}}
We obtained the perturbed parameter set $\Delta\theta = \theta_{0} + \theta_{p}$ from the group-level posterior mean values $\theta_{0}$ of \textit{ad libitum} fed rats given saline in the dark period. We used a perturbation $\theta_{p} = [0, -0.5, 0, 2, 1, 0, 0, 0]$ to perturb only the meal termination and feeding rate parameters. We took 100 sample trajectories from the model and averaged them in order to obtain Figure 4 E-I.

\end{document}